\documentstyle[preprint,prl,aps]{revtex}
\begin{document}
\draft
\title{Three-body recombination of ultra-cold atoms to a weakly
bound $s$ level}
\author{P.O. Fedichev$^{1,2}$, M.W. Reynolds$^{1}$,
and G.V. Shlyapnikov$^{1,2}$}
\address{{\em (1)} {Van der Waals - Zeeman Institute,
University of Amsterdam,\\
Valckenierstraat 65-67, 1018 XE Amsterdam, The Netherlands}\\
{\em (2)} {Russian Research Center Kurchatov Institute,\\
Kurchatov Square, 123182 Moscow, Russia}}
\date{March 5, 1996}
\maketitle
\begin{abstract}
We discuss three-body recombination of ultra-cold atoms
to a weakly bound
$s$ level. In this case, characterized by large and positive
scattering length $a$ for pair interaction, we find a repulsive
effective
potential for three-body collisions, which strongly reduces
the recombination probability and makes simple Jastrow-like
approaches absolutely inadequate. In the zero temperature limit
we obtain
a universal relation, independent of the detailed
shape of the interaction potential, for the (event) rate constant of
three-body recombination: $\alpha_{\rm rec}=3.9\hbar a^4/m$,
where $m$ is the atom mass.
\end{abstract}

\vspace{10mm}


Three-body recombination, the process in which two
atoms form a bound state and a third one carries away the binding
energy, is an important issue in the physics of
ultra-cold gases. This process represents the initial stage in the
formation of clusters intermediate in size between individual atoms
and bulk matter.
Three-body recombination limits achievable densities
in high-field-seeking spin-polarized atomic hydrogen
\cite{Kagan,VerhaarH} and in trapped alkali atom gases
(see \cite{Verhaar}
and references therein)
and, hence, places limitations on the possibilities to observe
Bose-Einstein condensation in these systems.

Extensive theoretical studies of three-body recombination in
ultra-cold hydrogen \cite{Kagan,VerhaarH} and
alkalis \cite{Verhaar}
showed that the rate constant of this process, $\alpha_{\rm rec}$,
strongly depends on the shape of the potential of
interaction between
atoms and on the energies of bound states in this potential.
In alkalis
the recombination is caused by elastic interatomic interaction,
and in
the zero temperature limit $\alpha_{\rm rec}$ varies
approximately as
$a^2$ \cite{Verhaar}, where $a$ is the scattering length for pair
interaction.

All these studies, except one in spin-polarized hydrogen
(see \cite{VerhaarH}), rely on Jastrow-like approximations
for the initial-state wavefunction of three colliding atoms.
Recent
progress in the quantum three-body problem for the case
where only zero orbital angular momenta of particle motion are
important \cite{Fedorov} opens a possibility for rigorous
calculations
of three-body recombination in ultra-cold atomic gases.
In this Letter we consider the extraordinary case of
recombination (induced by elastic interaction between atoms) to a
weakly bound $s$ level. The term ``weakly bound''
means that the size $l$ of the diatomic molecule in this
state is much larger than the characteristic radius of interaction
$R_e$ (the phase shift for $s$-wave scattering
comes from distances $r\alt R_e$).
In this case the scattering length is positive and
related to the binding energy $\varepsilon_0$ by
(see, e.g., \cite{LL})
\begin{equation}              \label{rel}
a=\hbar/\sqrt{m\varepsilon_0}\sim l\gg R_e
\end{equation}
($m$ is the atom mass), and elastic ($s$-wave) scattering
in pair collisions
is resonantly enhanced at collision energies $E\ll \varepsilon_0$.
As we show, large $a$ and $l$ imply a rather large recombination
rate constant $\alpha_{\rm rec}$.
At the same time, for large positive $a$ we find
a repulsive effective potential for three-body collisions, which
strongly reduces $\alpha_{\rm rec}$.
In the limit of ultra-low initial energies $E\ll \varepsilon_0$
we obtain a universal relation independent of the detailed
shape of the
interaction potential: $\alpha_{\rm rec}=3.9\hbar a^4/m$.

The dependence $\alpha_{\rm rec} \propto a^4$ can be understood
from qualitative arguments.
For atoms of equal mass the energy conservation law for the
recombination process reads
\begin{equation}               \label{law}
3\hbar^2k_f^2/4m=\varepsilon_0,
\end{equation}
where $k_f\sim 1/a$ is the final-state momentum of the
third atom relative to the center of mass of the molecule.
Recombination to a weakly bound $s$ level occurs in a
collision between two atoms, when a third atom is located inside a
sphere of radius $l\sim a$ around the colliding pair. For such
locations of the third atom, characterized by a statistical weight
$w\sim nl^3$ ($n$ is the gas density), this atom and one of the
colliding atoms form the weakly bound state with
probability of order unity.
The number of recombination events per unit time and unit volume,
$\nu_{\rm rec}=\alpha_{\rm rec}n^3$, can be estimated as $n^2\sigma
v(nl^3)$, where $\sigma=8\pi a^2$ is the cross section for pair
collisions. One may put velocity $v\sim \hbar k_f/m$,
which gives $\alpha_{\rm rec}\sim 8\pi\hbar a^4/m$.

One can also understand qualitatively the existence of a repulsive
effective
potential for three-body collisions and the reduction of
$\alpha_{\rm rec}$.
In the mean field picture the interaction in a
three-body system at (maximum of the three)
interparticle separations
$r\gg R_e$ can be written
as $4\pi \hbar^2n_{*}a/m$, where $n_{*}\sim 1/r^3$ is the
``particle density'' inside a sphere of radius $r$.
For $a>0$ this interaction is repulsive, which makes
the statistical
weight $w$ smaller than
$nl^3$ and decreases the numerical coefficient
in the above estimate
for $\alpha_{\rm rec}$.
The tail of the three-body effective potential
at $r\gg a$ was found in \cite{Zhen}. Arguments clearly
showing the absence of any ``kinematic'' repulsion
independent of the value and sign of $a$ are given
in \cite{Verhaar96}.

A particular system that should exhibit three-body recombination
to a
weakly bound $s$ level is a gas (or a beam) of helium
atoms. The He-He potential of interaction $V(r)$ has a well with a
depth
of $11$ K. There is only one
bound state in this well, with orbital angular momentum $j=0$ and
binding energy $\varepsilon_0\approx 1.3$ mK (see \cite{Tang} and
references therein).
The scattering length \mbox{$a\approx 100$ \AA} found for this
potential
satisfies criterion (\ref{rel}).
The existence of the ${\rm He}_2$ dimer, the world's largest
diatomic
molecule ($l\approx 50$ \AA), has been established experimentally
\cite{Toennies}.
Another system which is likely to have three-body
recombination to a
weakly bound $s$ level is spin-polarized
metastable triplet helium, a
gas of helium atoms in the $2^3S$ state with spins aligned.
The interaction potential \cite{Mey} for a pair of
spin-polarized He($2^3S$) atoms supports an $s$ level
with binding energy $\varepsilon_0\approx 2$ mK, which leads to
\mbox{$a\sim 100$ \AA} and
important consequences for the decay kinetics of this
system \cite{Gora94}.

We confine ourselves to three-body recombination of identical
atoms at collision energies $E\ll \varepsilon_0$ to a weakly bound
molecular $s$ level. In this case
the recombination rate constant $\alpha_{\rm rec}$ can be
found from the
equation
\begin{equation}            \label{rate2}
\nu_{\rm rec}\!=\!\alpha_{\rm rec}n^3\!=\!\frac{2\pi}{\hbar}\!\int
\!\!\frac{d^3k_f}{(2\pi)^3}\vert T_{if}\vert^2\delta
\!\left(\!\frac{3\hbar^2k_f^2}{4m}
-\varepsilon_0\!\right) \!\!\times \!\frac{n^3}{6}.
\end{equation}
Here $n^3/6$ stands for the number of triples in the gas,
$T_{if}=\int \psi_i\tilde V \psi_f^{(0)*}d^3xd^3x'$ is the
$T$-matrix element for three-body recombination, the coordinates
$({\bf x},{\bf x}')$ are specified in Fig.1,
$\psi_i$ is the true wavefunction of the initial
state of the triple, and $\psi_f^{(0)}$ is the wavefunction of free
motion of the third atom relative to the
center of mass of the molecule formed in the recombination event.
Both $\psi_i$ and $\psi_f^{(0)}$ can be written as a sum of three
components, each expressed in terms of one of the three
different sets of
coordinates (see Fig.1):
\begin{equation}             \label{psiisum}
\psi_i=\tilde \psi({\bf x},{\bf y})+\tilde \psi({\bf x}',{\bf y}')
+\tilde \psi({\bf x}'',{\bf y}''),
\end{equation}
\begin{eqnarray}           \label{psi0}
\psi_f^{(0)}=&&(1/\sqrt{3})[\phi({\bf x},{\bf y})+\phi({\bf
x}',{\bf y}')+\phi({\bf x}'',{\bf y}'')],  \\
&&\phi({\bf x},{\bf
y})\equiv\psi_0(x)\exp(i{\bf k}_f{\bf y}),  \nonumber
\end{eqnarray}
where $\psi_0$ is the wavefunction of the
weakly bound molecular state.
The interaction between colliding atoms is regarded as a sum
of pair interactions $V(r)$. The quantity $\tilde V$ is the part
of the interaction which is not involved in constructing
the wavefunction
(\ref{psi0}), i.e., if the molecule is formed by
atoms 1 and 2 (the first
term in Eq.\ref{psi0}), then
$\tilde V=V({\bf r}_1-{\bf r}_3)+V({\bf r}_2-
{\bf r}_3)$, et cet. Using Eq.(\ref{psi0}),
\begin{equation}           \label{T1}
\!T_{if}\!\!=\!2\!\sqrt{3}\!\!\int\!\!\!d^3\!xd^3\!x'\psi_0(x)
\!\cos\!\!\left( \!\frac{{\bf k}_f{\bf x}}{2}\!\!\right)
\!\!V(x')\!\exp(\!-i{\bf k}_f{\bf x}')\psi_i.\!\!
\end{equation}

The initial wavefunction of the triple is best
represented in hyperspherical
coordinates. The hyperradius, defined as
$\rho=(x^2/2+2y^2/3)^{1/2}$,
is invariant with respect to the transformations
$x,y\rightarrow x',y'
\rightarrow x'',y''$. The hyperangles are defined as
$\alpha=\arctan
(\sqrt{3}x/2y)$, and similarly for $\alpha'$ and $\alpha''$.
For $E\ll \varepsilon_0$ only zero orbital angular momenta of the
particle motion are important, and the wavefunction $\tilde\psi$
can be written as
\cite{Fedorov}
\begin{equation}        \label{psii}
\tilde\psi=\sum_{\lambda}\frac{F_{\lambda}(\rho)}{\sqrt{6}}
\frac{\Phi_{\lambda}
(\alpha ,\rho)}{\sin \alpha\cos\alpha}.
\end{equation}
The functions $\Phi_{\lambda}(\alpha,\rho)$
are determined by the equation
\begin{equation}              \label{Phieq}
\!-\!\frac{\partial^2\Phi_{\lambda}(\alpha,\rho)}
{\partial \alpha^2}\!+\!\!
\frac{2m}{\hbar^2}V(\!\sqrt{2}\rho\sin\alpha\!)\rho^2\!
\left( \!\Phi_{\lambda}(\alpha,\rho)\!\!+\!\frac{4}{\sqrt{3}}\!
\int_{\vert\pi/3-\alpha\vert}^{\pi/2-\vert\pi/6-\alpha\vert}\!\!
d\alpha'
\Phi_{\lambda}(\alpha',\rho)\!\!\right)\!\!=\!\!\lambda(\rho)
\Phi_{\lambda}(\alpha,\rho)\!,
\end{equation}
with boundary conditions $\Phi_{\lambda}(0,\rho)
=\Phi_{\lambda}(\pi/2,\rho)=0$
and normalization $\int_0^{\pi/2}\vert
\Phi_{\lambda}(\alpha,\rho)\vert^2d\alpha
=\pi/4$.
The sum in Eq.(\ref{psii}) is over all eigenvalues
$\lambda$ corresponding to
three free atoms at infinite interparticle separation.
At ultra-low
collision energies the lowest such $\lambda(\rho)$ alone
gives a very
good approximation, and we can confine ourselves to
this $\lambda$.
Then the function $F_{\lambda}(\rho)$ can be found from
the (hyper)radial
equation
in which the quantity $\lambda(\rho)$ serves as an
effective potential
\cite{Fedorov}.
Under the condition $E\ll \varepsilon_0$
at interparticle distances
much smaller than their De Broglie wavelength this
equation reads
\begin{equation}              \label{Feq}
\left(-\frac{\partial^2}{\partial\rho^2}-\frac{5}{\rho}
\frac{\partial}{\partial\rho}
+\frac{\lambda(\rho)-4}{\rho^2}\right)
F_{\lambda}(\rho)=0.
\end{equation}
The function $F_{\lambda}(\rho)$ should be finite for $\rho
\rightarrow 0$ and is normalized such that $F_{\lambda}(\rho)
\rightarrow 1$ for $\rho\rightarrow \infty$.

In our case the pair interaction potential $V(r)$
supports a weakly
bound $s$ level, and the scattering length is positive
and much larger
than the characteristic radius of interaction $R_e$ for
this potential.
For $\rho\gg R_e$ the function
$\Phi_{\lambda}(\alpha,\rho)$ takes
the form (cf. \cite{Fedorov})
\begin{equation}        \label{Phi}
\Phi_{\lambda}(\alpha,\rho)\!=\!\left\{
\begin{array}{l}
\!g(\rho)\alpha\!\left[ \left( \sqrt{2}\rho/a\right)
\!\sin(\pi\sqrt{\lambda}/2)
\chi_0(\sqrt{2}\rho\alpha)\!+\!(8/\sqrt{3})\!
\sin(\pi\sqrt{\lambda}/6)
\right]\!, \alpha\!<\!R_e/\rho\,\, \\
\!g(\rho)\sin\left[\sqrt{\lambda}
\left(\alpha-\pi /2\right)\right],
\alpha>R_e/\rho,
\end{array}\right.
\end{equation}
where
$g(\rho)=[1+\sin(\pi\sqrt{\lambda})/\pi\sqrt{\lambda}]^{-1/2}$
and $\chi_0(r)$ is the solution of the Schr\"odinger
equation for the
relative motion of a pair of particles,
\begin{equation}          \label{ShEq}
\left[-\frac{\hbar^2}{m}\left(\frac{\partial^2}{\partial r^2}+
\frac{2}{r}\frac{\partial}{\partial r}\right)
+V(r)\right]\chi_0(r)=0,
\end{equation}
normalized such that $\chi_0\rightarrow 1-a/r$ as
$r\rightarrow \infty$.
Matching the wavefunctions (\ref{Phi}) at
$\alpha=R_e/\rho\ll 1$, to zero order in $R_e/\rho$
we obtain the following
relation for $\lambda(\rho)$ at distances $\rho\gg R_e$
(cf. \cite{Fedorov}):
\begin{equation}         \label{matching}
\frac{\sqrt{2}\rho}{a}\sin\!\left(\!\sqrt{\lambda}
\frac{\pi}{2}\right)
\!+\!\frac{8}{\sqrt{3}}
\sin\!\left(\!\sqrt{\lambda}\frac{\pi}{6}\right)\!=
\!\sqrt{\lambda}\cos\!\left(\!\sqrt{\lambda}
\frac{\pi}{2}\right)\!.
\end{equation}
For $\rho\gg a$ this equation yields
$\lambda(\rho)=4+48a/\sqrt{2}\pi\rho$, and thus the
potential term in Eq.(\ref{Feq}) varies as $a/\rho^3$.
Eq.(\ref{matching}) is universal in the sense that
$\lambda$ depends only
on the ratio $\rho/a$, but not on the detailed shape of $V(r)$.
The same statement holds for $F_{\lambda}(\rho)$ at distances
$\rho\gg R_e$.

For infinite separation between particles, i.e., for
$\rho\rightarrow\infty$
and all hyperangles larger than $R_e/\rho$, we
have $\sqrt{\lambda}\approx 2$
and $\Phi_{\lambda}(\alpha,\rho)\approx \sin 2\alpha$.
Accordingly, from Eq.(\ref{psii})
with $F_{\lambda}(\rho)\rightarrow 1$, each $\tilde \psi$ in
Eq.(\ref{psiisum}) becomes equal to $\sqrt{2/3}$, and the initial
wavefunction $\psi_i \rightarrow\sqrt{6}$.

The ``effective potential'' $\lambda(\rho)$ and the
function $F_{\lambda}
(\rho)$ for three ground-state He atoms ($a\approx
100$\AA)
are presented in Fig.2 and Fig. 3. The potential $V(r)$
was taken from
\cite{Tang}. For $\rho\agt 100$\AA\ our numerically
calculated $\lambda(\rho)$ coincides (within 10\%) with
that following
from Eq.(\ref{matching}), ensuring a universal dependence of
$F_{\lambda}$ on $\rho/a$.
As $\lambda(\rho)$ is repulsive, $F_{\lambda}$ is
strongly attenuated
at $\rho\alt a$ (see Fig.3). This leads to a strong
reduction of $\psi_i$
when all three particles are within a sphere of radius $\sim a$.

We first consider the theoretical limit of weak binding,
where the scattering length $a$
and the binding energy $\varepsilon_0$ are related by
Eq.(\ref{rel}), the wavefunction of the bound molecular state
at distances $x\gg R_e$ is
\begin{equation}          \label{psib}
\psi_0(x)=\frac{1}{\sqrt{2\pi a}}\frac{1}{x}\exp
\left(-\frac{x}{a}\right),
\end{equation}
and the final momentum $k_f=2/\sqrt{3}a$. From
Eq.(\ref{psib}) one can see that the distance between
the two atoms which
will form the bound state should be of order $a$. To take away
the binding energy the third atom should approach one of
them to a distance
of order $R_e$. The main contribution to the integral in
Eq.(\ref{T1}) comes from distances $x\sim a$ and
$x'\sim R_e\ll a$. Therefore
we may put
$\rho\approx \sqrt{2/3}x$, $\alpha=\alpha''\approx\pi/3$,
and $\alpha'\approx
\sqrt{3}x'/2x$. Then the initial wavefunction takes the form
\begin{equation}        \label{ini}
\psi_i\approx (1/\sqrt{3})\chi_0(x')\tilde
F_{\lambda}(\sqrt{2}x/\sqrt{3}a),
\end{equation}
with $\tilde F_{\lambda}(z)\!=\!zF_{\lambda}(z)g(z)
\!\sin(\sqrt{\lambda(z)}\pi/2)$ and $z=\rho/a$.
Putting \mbox{${\bf k}_f{\bf x}'\!\!\approx\!\!0$} and using
\mbox{$\int\!\!d^3x'V(x')
\chi_0(x')\!=\!4\pi\hbar^2a/m$},
from Eq.(\ref{T1}) we
obtain \mbox{$T_{if}\!=\!48\pi^{3/2}\hbar^2a^{5/2}G/m$}, where
\begin{equation}             \label{G}
G=\int_0^{\infty}dz\sin(z/\sqrt{2})\exp(-z\sqrt{3/2})
\tilde F_{\lambda}(z).
\end{equation}
The main contribution to this integral
comes from $z\sim 1$ $(\rho\sim a)$, where $\lambda$ and
$F_{\lambda}$
(and, hence, $\tilde F_{\lambda}$) are universal functions of
$\rho/a$. Therefore $G$ is a universal
number independent of the potential $V(r)$. Direct calculation
yields $G=0.0364$.
With the above $T_{if}$ and $G$, from Eq.(\ref{rate2})
we arrive at the recombination rate constant
\begin{equation}                 \label{rate3}
\alpha_{\rm rec}=\frac{512\pi^2G^2}{\sqrt{3}}\frac{\hbar}{m}a^4
\approx 3.9\frac{\hbar}{m}a^4.
\end{equation}

The dependence $\alpha_{\rm rec}\propto a^4$, instead of
$\alpha_{\rm rec}\propto a^2$, is a consequence of the
recombination to a
weakly bound s level and can be also obtained within
the Jastrow approximation for the initial wavefunction:
\mbox{$\psi_{iJ}=\sqrt{6}\chi_0({\bf r}_1-{\bf r}_2)
\chi_0({\bf r}_2-{\bf r}_3)
\chi_0({\bf r}_3-{\bf r}_1)$}.
This approximation was proved to be a good approach for
atomic hydrogen \cite{VerhaarH} and was later used for
alkali atoms \cite{Verhaar}.
In our case, instead of Eq.(\ref{ini}),
we obtain $\psi_{iJ}\approx \sqrt{6}\chi_0(x')\chi_0^2(x)$
and arrive at Eq.(\ref{rate3}), with 4 orders of magnitude larger
numerical coefficient.
Such a very large discrepancy occurs because both
results are determined
by distances $x\sim a$, where in our (rigorous)
theory $\psi_i$ is
strongly reduced by the repulsive effective  potential
(see above). In the Jastrow approximation this reduction
is not present.
Moreover, $\psi_{iJ}$ is resonantly enhanced at distances $x<a$.
Thus, for large $a$ the Jastrow approximation gives a
wrong picture of
three-body collisions and is absolutely inadequate to describe
recombination to a weakly bound $s$ level.

The strong reduction of $\alpha_{\rm rec}$ due to the
presence of
a repulsive effective potential for three-body collisions can be
treated as ``quantum suppression'' of
three-body recombination (see related discussions in
\cite{Dalgarno,Verhaar96}). Nevertheless,
$\alpha_{\rm rec}$ remains
finite in the zero temperature limit. In fact,
due to large values
of $a$, it is rather large. It is also worth noting
that for large and {\it negative} scattering length the quantity
$\lambda(\rho)$ should have the form of a potential well, with a
repulsive core at small $\rho$, and the picture of recombination
collisions can be completely different.

In trapped gases the kinetic energy of the third atom acquired
in the recombination process
usually exceeds the trap barrier, and
such atoms escape from the trap.
Thus, the loss rate for atoms is
$\dot n=-Ln^3$, with $L=3\alpha_{\rm rec}$.
For three-body recombination of ground-state He
atoms Eq.(\ref{rate3})
gives $L\approx 2\cdot10^{-27}$ cm$^6$/s.
As the He-He interaction has $R_e\sim 15$\AA$\ll a$, this value of
$L$ is a very good approximation.
More accurate calculation, using $\lambda(\rho)$ and
$F_{\lambda}(\rho)$ determined for the He-He interaction (solid
curves in Fig.2 and Fig.3), gives a correction of 10\%.
The same $L$ is obtained for three-body recombination of
spin-polarized
He($2^3S$) atoms. In this case the result is less accurate,
since the
characteristic radius of interaction is somewhat larger
($R_e\sim 35$\AA).

Qualitatively, the picture of an effective repulsion in
three-body collisions, implying a strong reduction in the
recombination rate constant, can be valid for systems with
positive scattering length $a\sim R_e$.
One can find such systems among the ultra-cold alkali
atom gases.

We acknowledge stimulating discusions with J.T.M. Walraven.
The research of MWR is supported by
the Royal Netherlands Academy of Arts and Sciences.
This work was supported by the Dutch Foundation FOM,
by NWO through project NWO-047-003.036, by
INTAS and by the Russian Foundation for Basic Studies.

\begin{figure}[tbp]
\caption{Three possible sets of
coordinates for a three-body system.
The relative coordinates are ${\bf x}$, between two particles,
and ${\bf y}$,
between their center of mass and the third particle.}
\label{1}
\end{figure}

\begin{figure}[tbp]
\caption{ \protect
The ``effective potential'' $\lambda$ as a function of
$\rho/a$. The solid curve is obtained from Eq.(8) using
the ground state He-He potential ($a=100$\AA), and the dashed
from Eq.(12).
\label{2}
}
\end{figure}

\begin{figure}[tbp]
\caption{ \protect The wavefunction $F_{\lambda}(\rho/a)$
obtained from
Eq.(9). The solid curve corresponds to $\lambda(\rho)$ for the
ground state He-He potential, and the dashed curve to
$\lambda(\rho)$
from Eq.(12).
\label{3}
}
\end{figure}

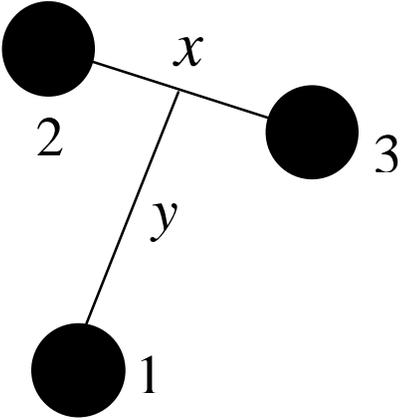
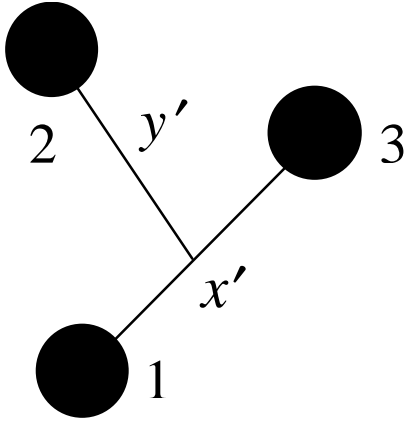
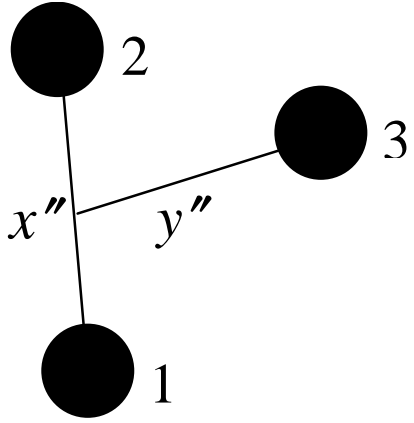
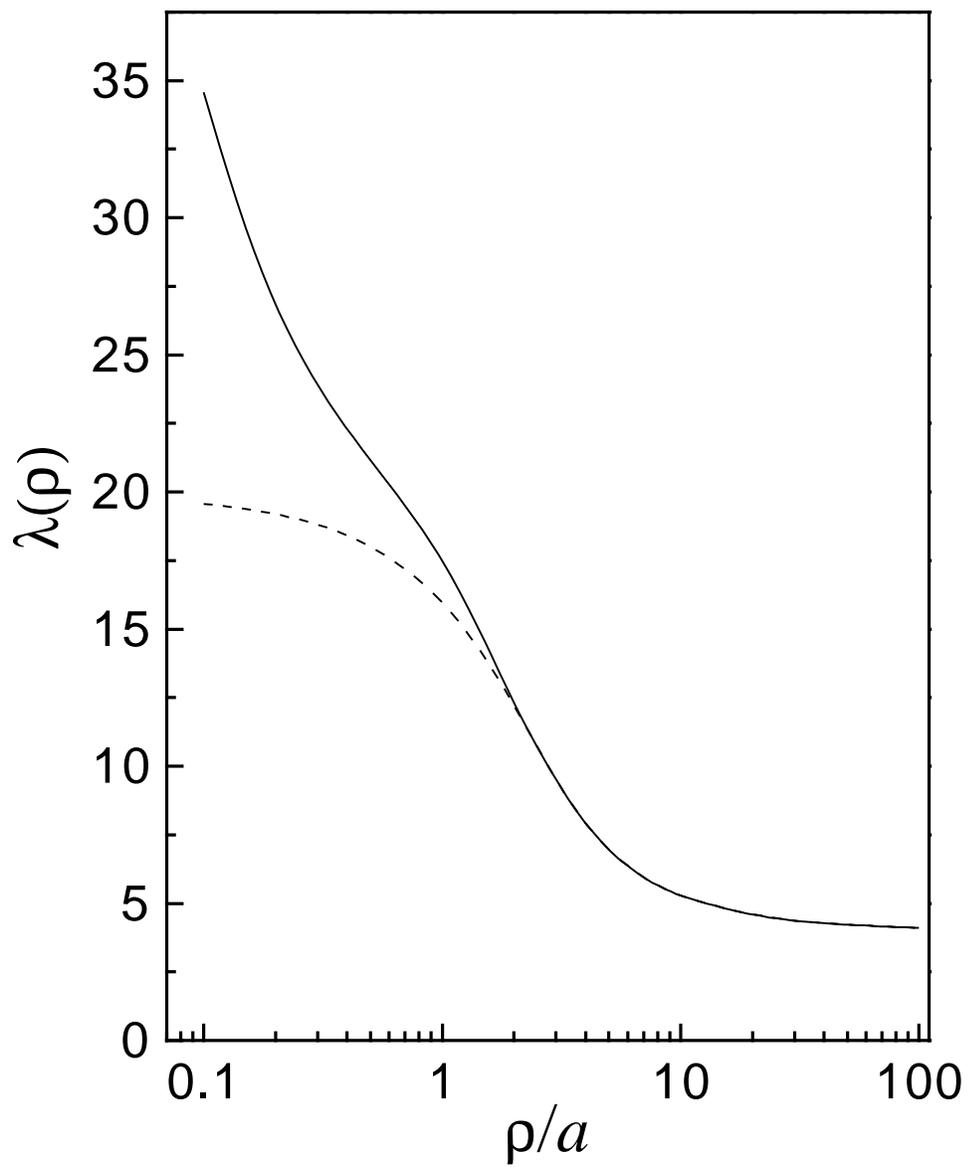
\end{document}